\documentstyle[twoside,aps,prl,multicol,epsf]{revtex} 
\begin{document}
\title{Bridge relations in Navier-Stokes turbulence}
\author{A. Celani $^{1,2}$ and D. Biskamp $^1$}
\address{$^1$Max Planck Institut f\"ur Plasmaphysik,
	D-85748 Garching, Germany}
\address{$^2$INFM Unit\`a di Torino Universit\`a, Italy} 
\maketitle
\begin{abstract}
The correlation between inertial range velocity fluctuations 
and energy dissipation in fully developed turbulence is studied
using high resolution direct numerical simulation.
Runs with microscale Reynolds number up to ${\cal R}_{\lambda} \simeq 300$
have been performed
in order to achieve the maximum extent of the scaling range compatible
with a proper resolution of the dissipative scales.
The bridge relations, a set of scaling relations
which express quantitatively the statistical dependence
of dissipation-scale velocity fluctuations 
on inertial-range ones, are verified.
Corrections to the bridge relations due to the finite size 
of the scaling range are investigated.

\end{abstract}

\begin{multicols}{2}
%\newpage

Fully developed Navier-Stokes turbulence is characterized by 
an energy flux from large scales, $L$, driven by some stirring
force, to small scales, $\eta$, where viscous dissipation is taking place.
Within the intermediate range of scales $L \gg r \gg \eta$, 
called the inertial range,
in which neither forcing nor dissipation
are active, longitudinal velocity differences
%$
%\bbox{w}(\bbox{x},\bbox{x}+\bbox{r})=
%\bbox{u}(\bbox{x}+\bbox{r})-\bbox{u}(\bbox{x})
%$
$
\delta_r u(x)=
(\bbox{u}(\bbox{x+r})-\bbox{u}(\bbox{x}))\cdot \bbox{r}/r 
$
display a scaling behavior: in homogeneous 
and isotropic turbulence, 
structure functions
%$
%S_p(r)=\langle |\bbox{u}(\bbox{x}+\bbox{r})-\bbox{u}(\bbox{x})|^p \rangle
%$
%\begin{equation}
%S_p(r)=\langle |\bbox{w}(\bbox{x},\bbox{x}+\bbox{r})|^p \rangle
%\label{1}
%\end{equation}
\begin{equation}
S_p(r)=\langle ( \delta_r u(x) )^p \rangle
\label{1}
\end{equation}
behave as power-laws $S_p(r)\sim r^{\zeta_p}$. 
According 
to the Kolmogorov 1941 theory $\zeta_p=p/3$, on the basis of dimensional 
arguments and self-similarity \cite{K41,MY}. Actually there is
strong evidence that 
exponents $\zeta_p$ are not those predicted by Kolmogorov,
but are a nonlinear function of $p$ (see e.g. \cite{Fri}).
This deviation from the Kolmogorov 1941 prediction is the most important 
signature of intermittency in the energy transfer process.

A basic consequence arising from the existence 
of an energy transfer directed to small scales is the 
 statistical dependence 
of small-scale velocity fluctuations on large-scale ones.
This dependence can be quantitatively investigated by looking at the 
behavior of inertial range two-scale 
velocity correlations such as 
\begin{equation}
F_{p,q}(r,R)=
\langle (\delta_r u(x))^p (\delta_R u(x))^q \rangle
%\langle |\bbox{w}(\bbox{x},\bbox{x}+\bbox{r})|^p 
%|\bbox{w}(\bbox{x},\bbox{x}+\bbox{R}|^q \rangle
\end{equation}
with $L \gg R \gg r \gg \eta$.
Under the usual scaling hypothesis and a mild universality assumption on
the small-scale structure of turbulence, it
can be shown that, to leading order, \cite{LP-1,LP-2}
\begin{equation}
F_{p,q}(r,R) \sim S_{p+q}(R)\frac{S_{p}(r)}{S_{p}(R)} \; 
\label{1b}
\end{equation} 
with an exception for $p=1$, where the prediction (\ref{1b})
does not apply due to the fact that $S_1(r)$ vanishes because of homogeneity
(see Refs. \cite{LP-2,BBT}).
These relations, and similar ones that have been derived in the
same context, have been successfully tested
in experiments \cite{FDLPS,BBT}. 
Since the scaling relations (\ref{1b}) hold exactly 
only when $L \gg R \gg r \gg \eta$,
their assessment
in numerical simulations is currently out of reach due to the limited
extent of the inertial range achieved with 
the actual computational capabilities.
It is worth noting
that the result (\ref{1b}) agrees with the prediction
of a multiplicative random process for the energy transfer 
mechanism \cite{BBT}.

Since the energy flux towards small scales will finally be
dissipated by viscous shear, it is reasonable to expect 
that, since this transfer induces correlations between
inertial-range fluctuations at different scales, it will also induce 
correlations between inertial-range fluctuations and dissipative-scale ones. 
This is the same kind of physical picture that lies at the base of
Kolmogorov's Refined Similarity Hypothesis \cite{K62}.

A recent result is that quantitative predictions
about the correlation between energy dissipation 
$\varepsilon({\bbox{x}}) \equiv \nu|\bbox{\nabla}\bbox{u}(\bbox{x})|^2$
and inertial range velocity fluctuations $\delta_r u(x)$
can be derived from Navier-Stokes equations
under the same hypotheses used in the derivation of (\ref{1b}) \cite{LP-3}.
These quantitative predictions involve correlations of the kind
\begin{eqnarray}
D^{(1)}_p(r) &=& \langle 
%\varepsilon(\bbox{x})|\bbox{w}(\bbox{x}+\bbox{r},\bbox{x})|^p 
\varepsilon(x) (\delta_r u(x))^p
\rangle \; , \label{2a}\\
D^{(2)}_p(r) &=& \langle 
%\varepsilon(\bbox{x})\varepsilon(\bbox{x+r})
%|\bbox{w}(\bbox{x}+\bbox{r},\bbox{x})|^p 
\varepsilon(x)\varepsilon(x+r)(\delta_r u (x))^p
\rangle \; , \label{2b}
\end{eqnarray}
which have been shown to follow the {\em bridge relations} \cite{LP-3}
\begin{eqnarray}
D^{(1)}_p(r)&\sim& \frac{S_{p+3}(r)}{r}\;,\label{3a} \\
D^{(2)}_p(r)&\sim& \frac{S_{p+6}(r)}{r^2} \; .
\label{3b}
\end{eqnarray}
The hierarchy of bridge relations extends to higher order correlations
involving three or more factors with energy dissipation.
Let us stress that in the presence of intermittency 
the scaling relations (\ref{3a},\ref{3b}) are different from the
scaling derived if dissipative scale fluctuations 
were independent of inertial scale ones.
Indeed the independence of $\varepsilon(x)$ from $\delta_r u$ would have led to
$D^{(1)}_p(r) \simeq \langle \varepsilon \rangle S_p(r) \sim r^{\zeta_p}$:
this is in contrast to (\ref{3a}), from which follows that 
$D^{(1)}_p(r) \sim r^{\zeta_{p+3}-1}$, since in presence of intermittency
we have that $\zeta_{p+3}-1 < \zeta_p$.

Note that the nature of scaling relations (\ref{1b}) --
which are derived only under general assumptions on scaling and universality 
of small-scale turbulence, and are shared by all systems displaying
a direct energy cascade --
is very different from the nature of the
bridge relations (\ref{3a},\ref{3b}), which are derived from the
Navier-Stokes equation and thus have a dynamical content.
It is noteworthy that the bridge relations coincide with the predictions
given by a multiplicative random process with a proper 
accounting of the viscous cutoff scales, and making explicitly
use of the Kolmogorov 4/5
law in the form $\zeta_3=1$ \cite{BBT}. The actual need for a relation like
Kolmogorov's law, which is itself a consequence of the Navier-Stokes equation,
to derive (\ref{3a},\ref{3b}) in the framework of multiplicative
models shows again the inherently dynamical nature of the bridge relations.

Among the bridge relations there are two renowned special cases.
The first one is (\ref{3a}) for $p=0$, 
which gives $\langle \varepsilon \rangle \sim S_3(r)/r$. 
This is the
Kolmogorov $4/5$ law \cite{Fri,MY}, which is the only bridge relation
that can be established within coefficients
%$
%S_3(r)/r=-4/5 \langle \varepsilon \rangle \; .
%$
\begin{equation}
\frac{S_3(r)}{r}=-\frac{4}{5} \langle \varepsilon \rangle \; .
\label{4}
\end{equation}
\noindent
The second well-known bridge relation is (\ref{3b}) for
$p=0$,  
%$
%\langle \varepsilon(\bbox{x})\varepsilon(\bbox{x+r}) \rangle \sim S_{6}(r)/r^2 
%$
$
\langle \varepsilon(x)\varepsilon(x+r) \rangle \sim S_{6}(r)/r^2 
$.
It implies the scaling for
the two-point dissipation correlation
%$
%\langle \varepsilon(\bbox{x})\varepsilon(\bbox{x+r}) \rangle \sim r^{-\mu}
%$ 
$
\langle \varepsilon(x)\varepsilon(x+r) \rangle \sim r^{-\mu}
$ 
with $\mu=2-\zeta_6$. This relation involving the intermittency exponent 
$\mu$ is the phenomenological 
``bridge relation'' \cite{Fri,MY}, 
hence the name chosen for (\ref{3a},\ref{3b}).

In this Letter we present the results of 
direct numerical simulations of fully developed homogeneous isotropic
turbulence which show clear evidence for the validity
of the bridge relations (\ref{3a},\ref{3b}) in the inertial range of scales.
 
The simulations solve the Navier-Stokes equation
on a grid with $512^3$ collocation points in a periodic cubic box of
linear size $2\pi$, using a standard pseudo-spectral method.
Time stepping is performed with a trapezoidal leap-frog scheme
and the dissipative term is integrated exactly. 
To obtain a statistically stationary state
the velocity field is forced for wavenumbers $k<2.5$. The forcing scheme
keeps the total energy constant in each of the first two shells
($0.5<k<1.5$ and $1.5<k<2.5$) with a ratio consistent with $k^{-5/3}$ 
\cite{CDKS}.
Two runs with different viscosity have been performed,
 with  microscale Reynolds numbers
${\cal R}_{\lambda} \simeq 300 $ and
${\cal R}_{\lambda} \simeq 220 $, 
 defined in the usual way as 
${\cal R}_{\lambda} \equiv v_0 \lambda/\nu$, where
$\lambda \equiv (15 \nu v_0^2 /\varepsilon)^{1/2}$
is the Taylor microscale 
and $v_0$ is the root mean square value of a single velocity component.
With the Kolmogorov length 
$\eta=(\nu^3/\varepsilon)^{1/4}$, the two runs correspond, respectively,
to $k_{max} \eta \simeq 1$ and $k_{max} \eta \simeq 2$, where $k_{max}=256$ 
is the largest wavenumber resolved. 
The run at the larger Reynolds number,
which has been chosen to maximize the scaling range,
resolves the dissipative scales
only marginally well, a fact that
might shed some doubt on the relevance of the results for the 
bridge relations (\ref{3a},\ref{3b}), since they involve the
energy dissipation $\varepsilon(x)$.  
We have therefore also performed the lower Reynolds number run, 
which has adequate resolution of the dissipative range,
and thus serves as a benchmark to
test the quality of the data obtained at ${\cal R}_{\lambda} \simeq 300$.
Ensemble averages are replaced by spatial averages and
time averages were performed over 
ten frames for each run in a time-span 
of approximately $7$ large eddy turnover times.

\begin{figure}[ht]
\epsfxsize=9.0truecm
\epsfbox{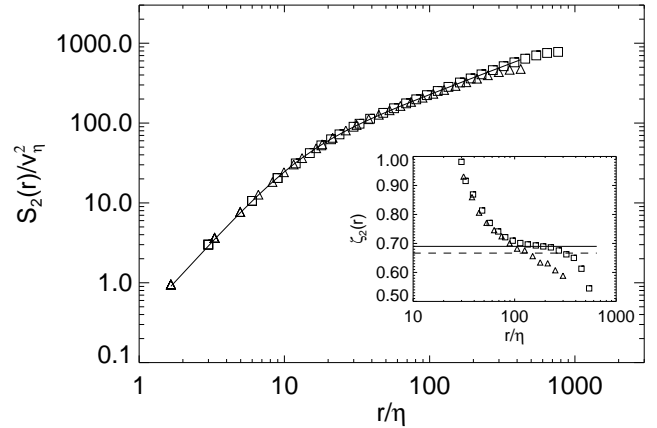}
\caption{Structure functions $S_2(r)/v_{\eta}^2$ for longitudinal velocity
increments at different Reynolds numbers, 
boxes ($\Box$) for ${\cal R}_{\lambda}\simeq 300$,
triangles ($\triangle$) for ${\cal R}_{\lambda} \simeq 220$. 
The solid line is the
Batchelor fit \protect(\ref{4b}) with $b=10.17$.
In the inset: the local slope $\zeta_2(r)$ 
for the data at ${\cal R}_{\lambda}\simeq 300$ ($\Box$)
and ${\cal R}_{\lambda} \simeq 220$ ($\triangle$), compared with the
value extracted by ESS $\zeta_2=0.69$ 
(solid line) and the Kolmogorov value $\zeta_2=2/3$ (dashed line)}
\label{fig:0}
\end{figure}

In Figure \ref{fig:0} is shown the second-order longitudinal
structure function $S_2(r)$ rescaled with respect 
to the square of the Kolmogorov velocity 
$v_{\eta} = \varepsilon^{1/3} \eta^{1/3}$, for the two runs at different
Reynolds number.

%%%%%%%%%%%%%%%%%%%%%%%%%%%%%%%%%%%%%%%%%%%%%%%%%%%%%%%%%%%%%%%%%%%%%%%%%%
\narrowtext
%%%%%%%%%%%%%%%%%%%%%%%%%%%%%%%%%%%%%%%%%%%%%%%%%%%%%%%%%%%%%%%%%%%%%%%%%%%
\noindent
According to the Batchelor parametrization, 
$S_2(r)$ can be expressed as \cite{MY,Batch}
\begin{equation} 
\frac{S_2(r)}{v^2_{\eta}}=
\frac{r^2/3 \eta^2}{ [1+(1+1/3b)^{3/2}(r/\eta)^2]^{1-\zeta_2/2} } \;,
\label{4b}
\end{equation}
which indeed fits the data very well, when the value $\zeta_2=0.69$
extracted by ESS technique is assumed \cite{ESS}.
The almost perfect agreement of the 
data for ${\cal R}_{\lambda} \simeq 220$
and ${\cal R}_{\lambda} \simeq 300$ 
at small and intermediate scales demonstrates the relevance
of results of the run at larger Reynolds number. 
In the inset of Figure \ref{fig:0} is also shown the local slope
of the second order structure function 
$\zeta_2(r)=d \ln S_2(r) /d \ln r$. Notwithstanding the limited
scaling range, for ${\cal R}_{\lambda} \simeq 300$ 
an interval in which $\zeta_2(r)$  is
approximately constant can be detected, with $\zeta_2(r) \simeq 0.69$,
definitely larger than the Kolmogorov value $2/3$ and in agreement
with the value obtained using ESS. 
In the case with ${\cal R}_{\lambda} \simeq 220$ no definite slope can be
inferred from $\zeta_2(r)$, and one has to rely completely upon ESS
to determine the scaling exponent.

\begin{figure}[ht]
%\centerline{\epsfxsize=220pt\epsfysize=183.68pt\epsfbox{fig1.eps}}
\epsfxsize=9.0truecm
\epsfbox{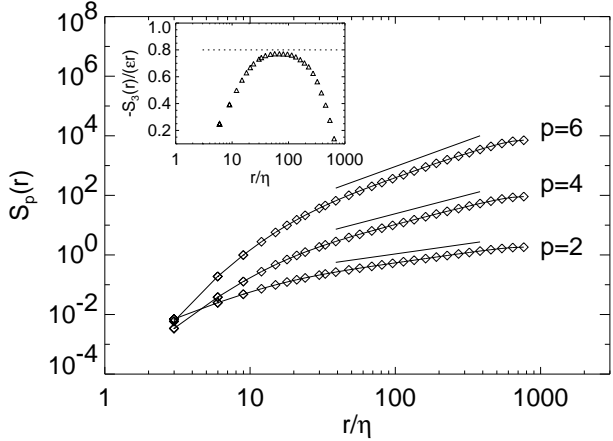}
\caption{Structure functions $S_p(r)$ for longitudinal velocity
increments with $p=2,4,6$. Solid lines are power laws with exponents
$\zeta_2=0.69,\zeta_4=1.27,\zeta_6=1.77$ extracted by ESS technique.
Data are shifted on the vertical axis for viewing purposes
(factor $10^2$ for $p=4$ and $10^4$ for $p=6$).
In the inset is shown the Kolmogorov $4/5$ law \protect(\ref{4}).
Note the linear scale on the vertical axis.}
\label{fig:1}
\end{figure}

In Figure~\ref{fig:1} we show the longitudinal structure functions for the
lowest even orders, and a check of Kolmogorov $4/5$ law (\ref{4}), 
both for ${\cal R}_{\lambda} \simeq 300$.

%%%%%%%%%%%%%%%%%%%%%%%%%%%%%%%%%%%%%%%%%%%%%%%%%%%%%%%%%%%%%%%%%%%%%%%%%
\narrowtext
%%%%%%%%%%%%%%%%%%%%%%%%%%%%%%%%%%%%%%%%%%%%%%%%%%%%%%%%%%%%%%%%%%%%%%%%%%%
\noindent
These results are in agreement with previous numerical simulations \cite{CCS}.

\begin{figure}[hb]
%\centerline{\epsfxsize=220pt\epsfysize=183.68pt\epsfbox{fig1.eps}}
\epsfxsize=9.0truecm
\epsfbox{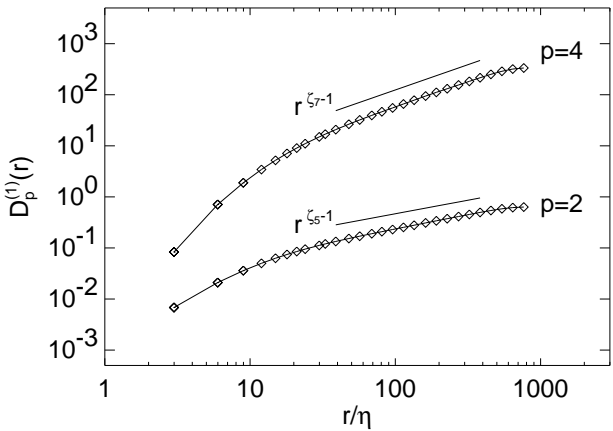}
\caption{Correlations $D^{(1)}_p(r)$ for longitudinal velocity
increments with $p=2,4$. Solid lines are power laws with exponents
$\zeta_5-1=0.53,\zeta_7-1=0.99$, which are the scaling exponents
predicted by the bridge relations (\ref{3a}). 
The exponents $\zeta_p$ have been obtained
 by ESS technique.}
\label{fig:2}
\end{figure}

In Figure~\ref{fig:2} we examine the
scaling behavior of the correlations
$D^{(1)}_p(r)=\langle \varepsilon(x) (\delta_r u(x))^p \rangle$.

%%%%%%%%%%%%%%%%%%%%%%%%%%%%%%%%%%%%%%%%%%%%%%%%%%%%%%%%%%%%%%%%%%%%%%%%%
\narrowtext
%%%%%%%%%%%%%%%%%%%%%%%%%%%%%%%%%%%%%%%%%%%%%%%%%%%%%%%%%%%%%%%%%%%%%%%%%%%
\noindent
According to the bridge relations (\ref{3a}), the scaling
$D^{(1)}_p(r) \sim r^{\zeta_{p+3}-1}$ should hold within the inertial
range. The data are in good agreement
with the theoretical prediction 
within the accuracy allowed by the limited scaling range.

\begin{figure}[ht]
\epsfxsize=9.0truecm
\epsfbox{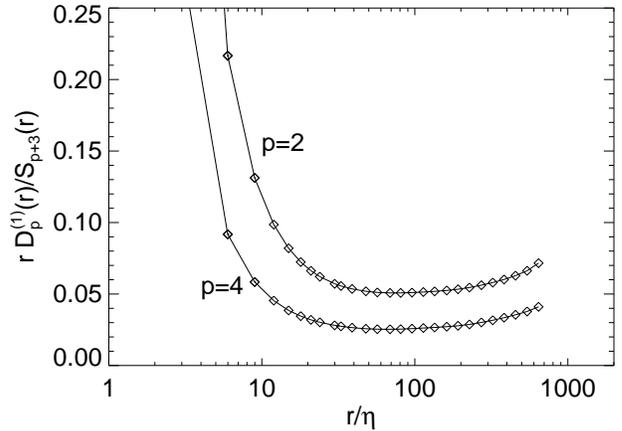}
\caption{The ratio $r D^{(1)}_p(r)/ S_{p+3}(r)$, 
approximately independent of $r$ for 
intermediate values of $r$, verifying the bridge relations 
(\protect\ref{3a}).
Odd order $S_p(r)$ are computed taking the absolute value of 
velocity differences.} 
\label{fig:3}
\end{figure}

In order to have a more accurate assessment of the bridge relations (\ref{3a}), 
it is useful to display the behavior of the ratio
$r D^{(1)}_p(r)/S_{p+3}(r)$, which should be independent of $r$
within the inertial range. As shown in Figure~\ref{fig:3}
there is a clear plateau (note the linear scale on the
vertical axis) in an interval of values of
$r$ coincident with the scaling range of the even-order
structure functions, confirming the results of Figure~\ref{fig:2}.  

\begin{figure}[ht]
\epsfxsize=9.0truecm
\epsfbox{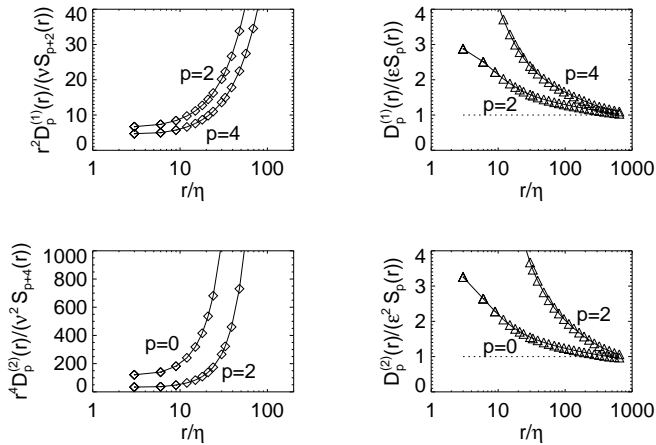}
\caption{Upper Left: The ratio $r^2 D^{(1)}_p(r)/(\nu S_{p+2}(r))$
becomes independent of $r$ for $r/\eta \rightarrow 0$, according to (\ref{5}).
Upper Right: The asymptotic decorrelation of energy dissipation
and velocity differences reflects in the behavior of
$D^{(1)}_p(r)/(\varepsilon S_p(r))$ which tends to unity for $r/L \sim 1$.
Lower Left: $r^4 D^{(2)}_p(r)/ (\nu^2 S_{p+4}(r))$
becomes independent of $r$ for $r \rightarrow \eta$, according to (\ref{7a}).
Lower Right: Asymptotic behavior of
$D^{(2)}_p(r)/(\varepsilon^2 S_p(r))$}
\label{fig:8}
\end{figure}

%%%%%%%%%%%%%%%%%%%%%%%%%%%%%%%%%%%%%%%%%%%%%%%%%%%%%%%%%%%%%%%%%%%%%%%%%
\narrowtext
%%%%%%%%%%%%%%%%%%%%%%%%%%%%%%%%%%%%%%%%%%%%%%%%%%%%%%%%%%%%%%%%%%%%%%%%%%%
\noindent
Figure~\ref{fig:3} also shows that the bridge relations break down at
small and at large $r$. 
Indeed close to and below the dissipative scale 
$\eta$ we have the smooth behavior 
$ 
D^{(1)}_p(r) \sim \langle \varepsilon^{p/2+1} \rangle r^{p}/ \nu^{p/2}
$ 
and 
$
S_{p}(r) \sim \langle \varepsilon^{p/2} \rangle r^{p}/ \nu^{p/2}
$ 
which leads to the scaling relation 
\begin{equation}
D^{(1)}_p(r) \sim \nu S_{p+2}(r)/r^2 \qquad \mbox{for} 
\qquad r/\eta \rightarrow 0 \; .
\label{5}
\end{equation} 
Furthermore, when the distance $r$ becomes comparable with 
the integral scale $L$ we expect that the statistics of velocity differences
becomes independent of the dissipation.
Thus the correlation $D^{(1)}_p(r)$ must asymptotically become
\begin{equation}
D^{(1)}_p(r) \simeq \langle \varepsilon \rangle S_{p}(r) \qquad \mbox{for} \qquad r > L \; .
\label{6}
\end{equation}
The asymptotic behavior for $D^{(1)}_p(r)$ is shown in the upper 
panels of Figure \ref{fig:8}.

\begin{figure}[ht]
\epsfxsize=9.0truecm
\epsfbox{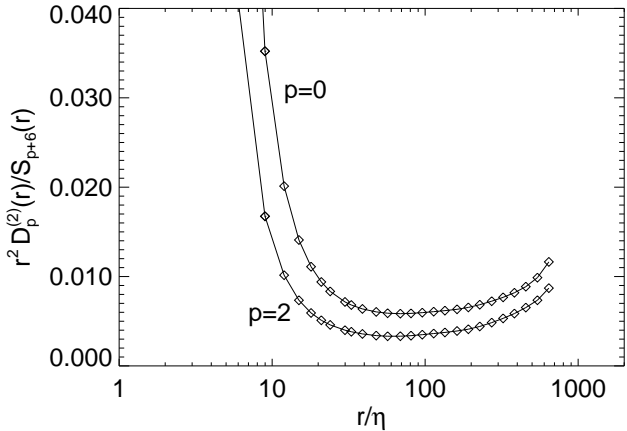}
\caption{The ratio $r^2 D^{(2)}_p(r)/ S_{p+6}(r)$,
approximately independent of $r$ for
intermediate values of $r$, verifying the bridge relations 
(\protect\ref{3b}).}
\label{fig:5}
\end{figure}

In order to check the validity of the bridge relations (\ref{3b})
we plotted the ratio $r^2 D^{(2)}_p(r)/ S_{p+6}(r)$ versus $r$, which 
-- in agreement with (\ref{3b}) --
 is constant for values of $r$ within the inertial 
range of scales, as shown in Figure~\ref{fig:5}. 
The case $p=0$ corresponds to the phenomenological bridge relation
$\langle \varepsilon(x)\varepsilon(x+r) \rangle \sim r^{-\mu} 
\sim r^{2-\zeta_6}$, with $\mu=0.22$. The agreement is slightly
less good than that for $D^{(1)}_p(r)$, an effect that can be ascribed
to the higher order of the structure functions involved. 

%%%%%%%%%%%%%%%%%%%%%%%%%%%%%%%%%%%%%%%%%%%%%%%%%%%%%%%%%%%%%%%%%%%%%%%%%
\narrowtext
%%%%%%%%%%%%%%%%%%%%%%%%%%%%%%%%%%%%%%%%%%%%%%%%%%%%%%%%%%%%%%%%%%%%%%%%%%%
\noindent
For $D^{(2)}_p(r)$ there are also departures from the bridge relations 
approaching
either the dissipative scale $\eta$ or the integral scale $L$,
with leading scaling laws 
\begin{eqnarray}
D^{(2)}_p(r) \sim \nu^2 S_{p+4}(r)/r^4 \qquad \mbox{for} 
\qquad r/\eta \rightarrow 0 
\label{7a} \\
D^{(2)}_p(r) \simeq  \langle \varepsilon \rangle^2 S_{p}(r) \qquad \mbox{for} \qquad r > L \; ,
\label{7b}
\end{eqnarray}
which follow respectively from the smoothness of the velocity field
at dissipative scales and from the independence of the dissipative range
 on the forcing scales, and are shown in the lower panels of Figure
\ref{fig:8}.

%%%%%%%%%%%%%%%%%%%%%%%%%%%%%%%%%%%%%%%%%%%%%%%%%%%%%%%%%%%%%%%%%%%%%%%%%
\narrowtext
%%%%%%%%%%%%%%%%%%%%%%%%%%%%%%%%%%%%%%%%%%%%%%%%%%%%%%%%%%%%%%%%%%%%%%%%%%%

In summary,
the physical picture emerging is that the straining process
transforms inertial scale fluctuations into dissipation scale 
fluctuations, thus inducing a statistical dependence
of the latter on the former with an inherently dynamical nature.
The ``bridge relations'' are a quantitative expression of this dependence,
which determine the scaling exponents of correlations
between dissipation and inertial range velocity fluctuations
in terms of the usual scaling exponents of structure functions.
These relations have been successfully tested in the framework
of direct numerical simulations of Navier-Stokes turbulence.
The analysis has been performed with microscale Reynolds number up to 
${\cal R}_{\lambda} \simeq 300$, a value that allows to achieve the maximum
extent of the scaling range without giving up the
proper resolution of the dissipative scales.
Experimental confirmation of the results here presented would be 
heartily welcome.

We thank G. Boffetta and L. Biferale for useful discussions and suggestions.
Numerical simulations were performed at the Max Planck Institute
f\"ur Plasmaphysik, using a CRAY T3E parallel computer.
A. Zeiler is acknowledged for the development of the skeleton 
of the numerical code.
The help of J. Pichlmeier from Cray for assistance in 
high performance computing is gratefully acknowledged.
A.C. has been supported by the European Community TMR grant ERBFMICT-972459,
and by INFM (PRA TURBO).

\end{multicols}
\end{document}